\documentstyle[aps,prl,multicol,epsf]{revtex}

\def\({\begin{equation}}
\def\){\end{equation}}
\begin{document}                
%
%

\title{Low temperature transport in the XXZ model}
\author{B. N. Narozhny$^{1}$, A. J. Millis$^2$, N. Andrei$^{1}$}
\address{$^1$Department of Physics and Astronomy Rutgers University, Piscataway, NJ, 08855. \\
$^2$ Department of Physics and Astronomy, The Johns Hopkins University, 3400 North 
Charles St., Baltimore, MD 21218. }
\maketitle
\begin{abstract} 
We present evidence suggesting that spin transport in the gapless phase of the 
S=1/2 XXZ model is ballistic rather than diffusive. We map the model onto a 
spinless fermion model whose charge stiffness determines the spin transport of 
the original model. By means of exact numerical diagonalisation and finite 
size scaling we study both the stiffness and the level statistics. We show 
that the stiffness is non-zero at low temperatures so that the transport is 
ballistic. Our results suggest that the non-zero stiffness is due to the fact 
that even in the presence of Umklapp scattering a non-zero fraction of states 
remain degenerate in the thermodynamic limit.
\end{abstract}

\begin{multicols}{2}

The problem of transport in a non-disordered interacting many particle system
is one of the oldest unsolved problems in solid state physics. A particular
case which has attracted recent attention is spin diffusion in one dimensional
spin systems. Recent experiments indicated that S=1 chains with the gap in the
excitation spectrum display diffusive behavior \cite{tak1}, in reasonable 
agreement with theoretical work \cite{sach} relating diffusion to classical 
scattering of excitations near the gap edge. On the other hand, measurements
on gapless S=1/2 chains \cite{tak2} show a different behavior. The authors 
fit their data with a diffusion constant which is much larger than either the 
value found experimentally for the S=1 chains or the value 
${\cal D}_s \sim J\sqrt{2\pi S(S+1)/3}$ expected from classical considerations
\cite{whit}. We believe that the measured value for ${\cal D}_s$ in the S=1/2
system is so large that it implies that the diffusion is not an intrinsic
property of an S=1/2 spin system but is due to a weak coupling to other degrees 
of freedom (for example to phonons \cite{meme}).

A theoretical analysis based on a continuum limit Luttinger liquid representation 
\cite{mill} suggested that the diffusion constant was associated with Umklapp 
operators and was finite but exponentially large in the Umklapp gap. Damle and
Sachdev \cite{sach} provided a detailed analysis indicating diffusive behavior 
for the S=1 chain, however, their work suggests that in general low energy 
excitations of gapped systems display diffusive behavior even if the effective 
low energy theory describing these excitations is integrable. On the contrary, 
Zotos and collaborators have argued that integrable models exhibit ballistic 
transport while non-integrable models are diffusive \cite{zot2}. Recently 
Fabricius and McCoy \cite{mcco} have shown that numerical computations of the 
long time behavior of the correlation functions of the S=1/2 XXZ chain in the 
$T=\infty$ limit are consistent with ballistic transport if the model has a XY 
anisotropy but suggest that at the isotropic Heisenberg point the transport is 
not ballistic. Very recently, Monte Carlo\cite{kirc} and 
Bethe-ansatz \cite{pere}analyses of the
stiffness of related models have appeared, reporting similar conclusions.

In this paper we approach the question in a different way. We use exact numerical
diagonalisation and finite size scaling to study the spin stiffness of the XXZ
model and the behavior of the energy levels which leads to a non-zero stiffness.

The XXZ model is defined by the Hamiltonian

\begin{equation}
\hat H_{XXZ} = {J}\sum\limits_{n} (S^x_i S^x_{i+1} + S^y_i S^y_{i+1}
+ \Delta S^z_i S^z_{i+1}).
\label{heis}
\end{equation}

\noindent
At $T=0$ the gapless phase $-1 < \Delta \le 1$ of this model is characterized by 
a non-zero spin stiffness ${\cal D}_s$\cite{shas} (in the gapped phase $|\Delta|>1$
the stiffness is zero). The question of interest here is whether at $T>0$ the
stiffness ${\cal D}_s$ remains non-zero, implying ballistic transport, or 
${\cal D}_s$ vanishes, implying non-ballistic (and perhaps diffusive) transport.

The XXZ model is equivalent via the 
Jordan-Wigner transformation to the spinless fermion model,

\begin{eqnarray}
\hat H = && J\sum\limits_{n} \left [ -{1\over{2}}(c_{n}^\dagger c_{n+1}
+ h.c.) \right. \nonumber\\
&& 
\nonumber\\
&& + \left. 
\Delta\left (c_{n}^\dagger c_{n} - {1\over{2}}\right )
\left (c_{n+1}^\dagger c_{n+1} - {1\over{2}}\right )\right ].
\label{ferm}
\end{eqnarray} 

\noindent
In this mapping the fermion density-density
correlation function represents the $S^z-S^z$ correlator and therefore the description
of fermion transport directly translates into the spin
language. In particular the real part of the frequency dependent conductivity
may be written as

\begin{eqnarray}
{\cal R}e \;  \sigma (\omega ) = 2 \pi D_c \delta (\omega ) + \sigma_{reg} (\omega ),
\end{eqnarray}

\noindent
defining the charge stiffness $D_c$. If $D_c \ne 0$ the model has infinite conductivity
whereas if $D_c=0$ one has either a normal conductor 
(${D_c=0,} \; \sigma_{reg}(\omega \rightarrow 0)>0$) with diffusive transport 
or an ideal insulator (${D_c=0,} \; \sigma_{reg}(\omega \rightarrow 0)=0$). In the spin 
language $D_c$ becomes the spin stiffness, which, if non-zero, corresponds to ballistic 
spin transport, whereas if it is equal to zero then the long time relaxation at finite
temperatures is expected to be diffusive. 

As was first noted by Kohn \cite{kohn}, in systems with periodic boundary conditions 
the stiffness at $T=0$ can be related to the response of the ground state energy $E_0$ 
to a magnetic flux $\phi$, which modifies the hopping term in the Hamiltonian 
Eq.\ (\ref{ferm}) by the usual Peierls phase factor $t\rightarrow t\exp(\pm\imath\phi/L)$,  
so that $D_c= (L/2)\partial^2E_0 / \partial \phi^2 (\phi\rightarrow 0)$, where $L$ is the 
system size. This phase factor can be absorbed into twisted boundary conditions for the 
wave functions.

Kohn's method has been recently generalized to finite temperatures \cite{zot2}, 

\begin{equation}
D_c = {\displaystyle{L \over{2{\cal Z}}}} \sum\limits_{n} 
{\displaystyle{1 \over {2}}} \displaystyle{ {\partial^2 E_n} \over {\partial \phi^2}}
\displaystyle{e^{-\beta E_n}} 
\label{D}
\end{equation}

\noindent
where ${\cal Z}$ is the partition function of the system and $n$ 
labels exact eigenstates.

We further rewrite Eq.\ (\ref{D}) as $D_c=D_1+D_2$, with 

\begin{equation}
D_1 = - {\displaystyle{L \over {2\beta}}} {\displaystyle{1 \over{\cal Z}}}
\displaystyle{ {\partial^2 {\cal Z}} \over {\partial \phi^2}}
\label{D1}
\end{equation}

\noindent
and 

\begin{equation}
D_2 = {\displaystyle{{\beta L} \over {2}}} {\displaystyle{1 \over{\cal Z}}}
\sum\limits_{n}  \left(\displaystyle{ {\partial E_n} \over {\partial \phi}}\right)^2
\displaystyle{e^{-\beta E_n}}.
\label{D2}
\end{equation}

\noindent
The advantage of this representation is that it separates $D_c$ into a thermodynamic part 
(depending only on derivatives of ${\cal Z}$) and a {\it positive} part, depending on 
current-carrying ($j_{nn}\sim\partial E_n/\partial \phi$) states. The thermodynamic part
may be seen to give no contribution to the charge stiffness at $T>0$ \cite{gome}; 
for example at low temperatures $T \ll J$ 

\begin{equation}
D_1(T\gg 2\pi/vRL) = L T \exp(-{{2\pi L T}\over{ v R}}).
\label{d11}
\end{equation}

\noindent
with $R^2 = (1-(1/\pi)\cos^{-1}\Delta)/2\pi$.
Thus at $T>0$ any non-zero $D_c$ must be due to $D_2$ which essentially counts the number 
of thermally accessible current carrying states. 

Time reversal invariance implies that for a non-degenerate state 
$\partial E_n/\partial\phi (\phi\rightarrow 0) = 0$. Current carrying states occur in 
degenerate pairs which 
which are split by the application of magnetic flux. A sufficient condition for a 
non-vanishing $D_c$ is to have a non-zero fraction of current carrying states with 
$\partial E_n/\partial\phi \sim 1/L^{1/2}$. In what follows, we investigate $D_c$ and the 
statistics of the current carrying states numerically.

To compute the eigenvalues we notice that for any finite size chain the Heisenberg 
Hamiltonian Eq.\ (\ref{heis}) is just a hermitian matrix.
Our strategy is to construct this matrix for $\phi= {1\times 10^{-4}},{2\times 10^{-4}}$, 
${3\times 10^{-4}}, {4\times 10^{-4}}$ and use the
exact diagonalisation to obtain the eigenvalues and compute the derivatives. We use the 
standard QL routine from the Numerical Recipes package \cite{numr} which calculates the 
eigenvalues through a series of orthogonal transformations with accuracy given by 
machine precision. Our choices of $\phi$ lead to $\sim 10^{-6}$ accuracy for the 
derivative values. The size of matrices that could be diagonalised by the routine is
limited by computer memory; for a $N\times N$ matrix it requires $\approx 8 N^2$ bytes of
storage space. With the availiable computer memory of about 360 MB we can diagonalise 
matrices up to $N=7000$ which limited us to chain sizes $L\le 14$. 

\begin{figure}
\centerline{\epsfxsize=5.1cm \epsffile{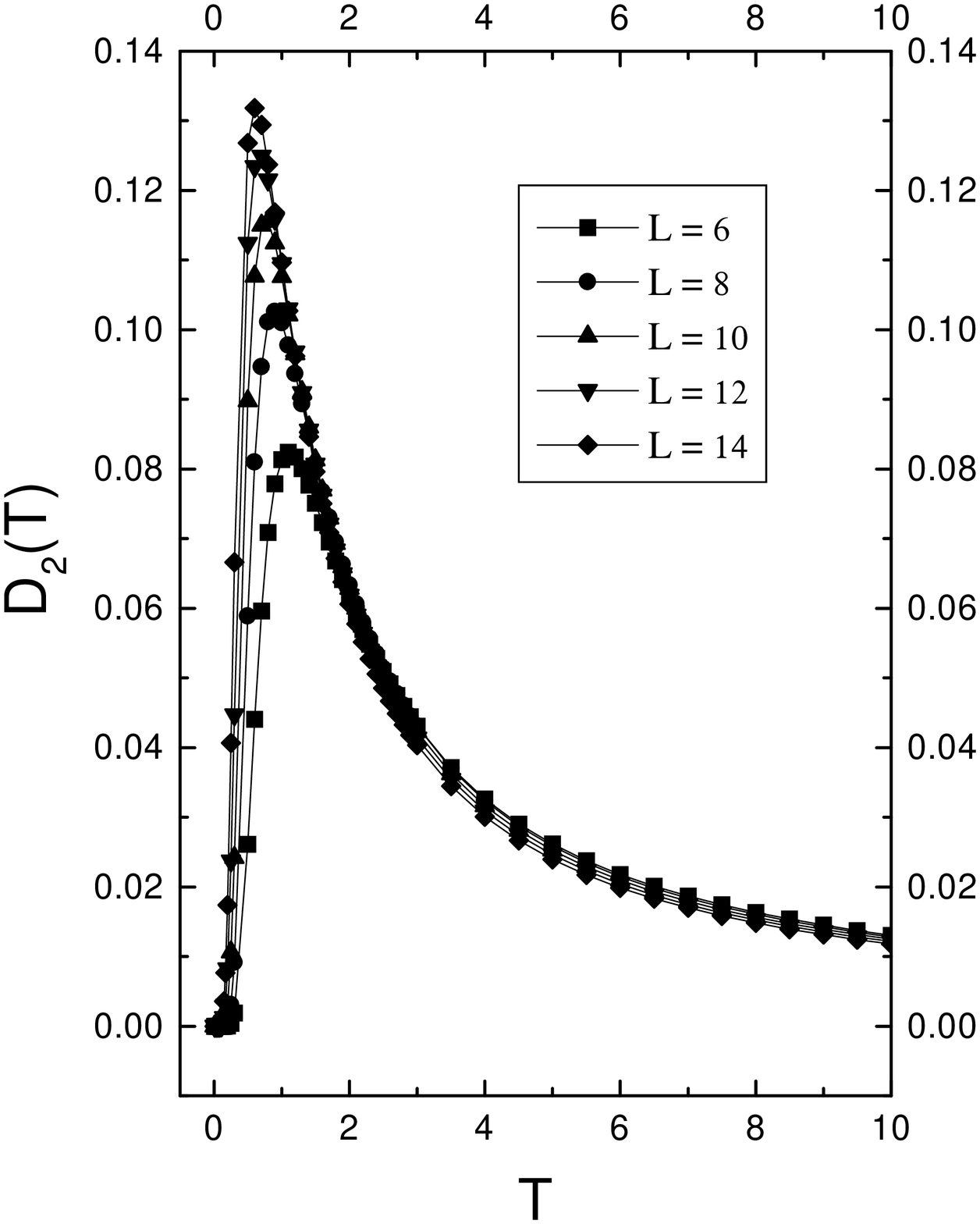}}
Fig. 1.
{$D_2 (T)$ for different system sizes for $\Delta = 0.4$.} 
\label{fig421}
\end{figure}

The result of the calculation is presented in Fig. 1. For all system sizes we found
$D_2$ to be non-zero. At small temperatures the value of $D_2$ appears to grow with system 
size (especially the peak value). 
At large temperatures all eigenvalues become involved in the sum 
Eq.\ (\ref{D2}) and the temperature dependence is defined by the prefactor $1/T$, 
while the value of $D_2$ decreases with the system size. 

To investigate the finite size scaling in Fig. 2 we plot
$(D_2 T)_\infty =\lim\limits_{T\rightarrow\infty} D_2 T$ 
versus the inverse system size for different values of the interaction.
The symbols represent the actual data points and the best fit lines are continued to the
infinite size ($1/L=0$). We are unaware of theoretical results for the large $L$ 
behavior of $(D_2 T)_\infty$; our numerical results are consistent with the ansatz 
$(D_2 T)_\infty (L) = A + B/L + ... $ with $A, B$ depending on the interaction, but with 
$A$ always positive for $\Delta \le 1$.
For small $\Delta$ the best fit line is flat and the fit using four largest sizes is
excellent (error estimate for parameter $A$ is $0.7\%$). For $\Delta =0.6$ we find 
$A\approx 0.076$, $B\approx 0.32$ with $3.6\%$ error. At the isotropic point $\Delta = 1$
the best straight line fit yields $A\approx 0.029$, $B\approx 0.46$ but with rather
larger $11\%$ error leading us to question whether we have assumed the correct functional 
form. We note, however, that fits to the form $(D_2 T)_\infty (L) = C/L^\theta$ lead to 
even larger errors, so the hypothesis  $(D_2 T) (L\rightarrow\infty) \rightarrow 0$ is
inconsistent with our data.

\begin{figure}
\centerline{\epsfxsize=5.3cm \epsffile{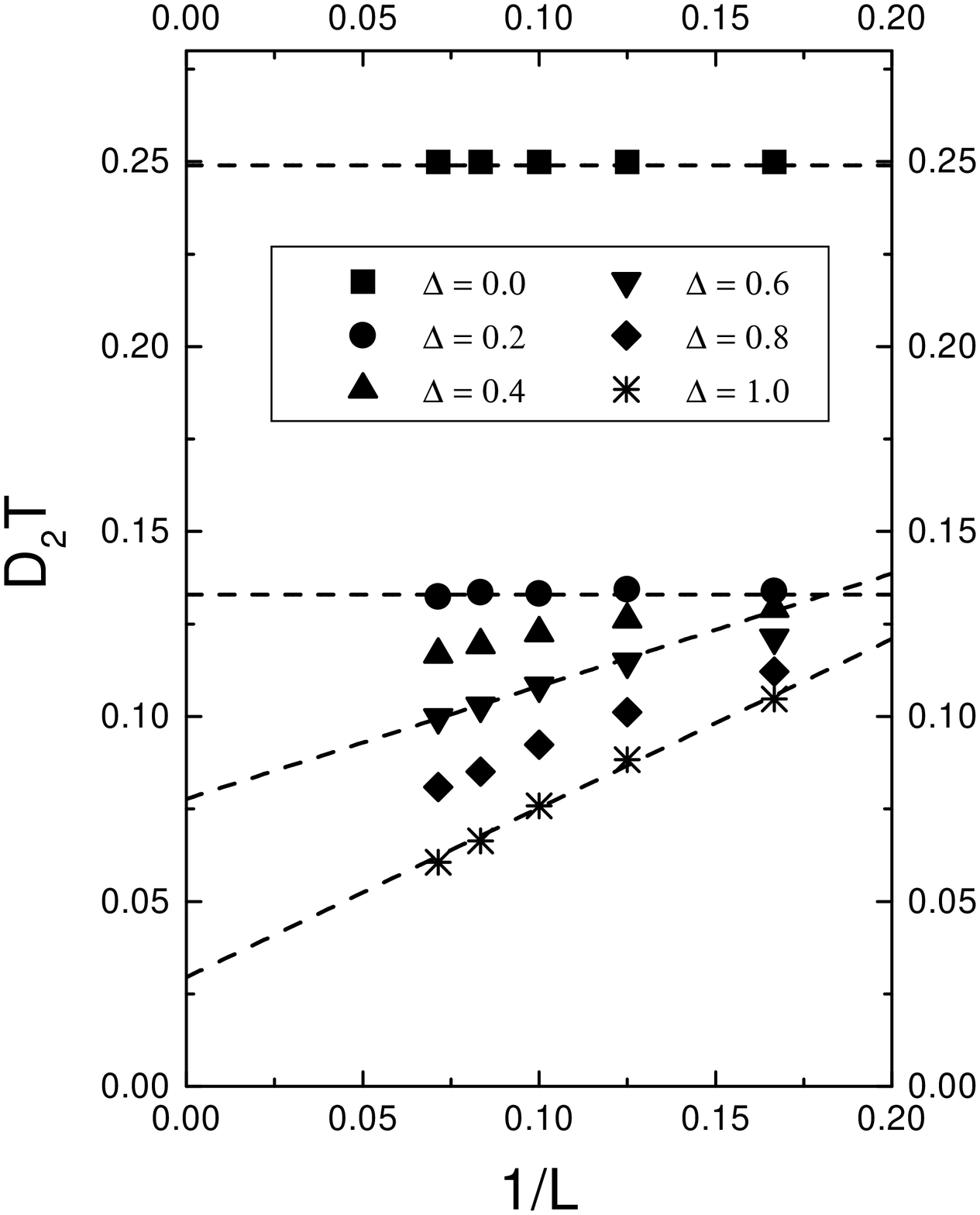}} 
Fig. 2. 
{$(D_2 T)$  plotted against inverse system size at $T=50$.}
\label{fig422}
\end{figure}

The scaling of $D_2$ can be expressed in terms of the size dependence of the current 
carried by a typical excited state.  
For the free case a typical state contains a total number $P \sim L$ of fermions 
excited above both left and right Fermi points and has a $\sqrt{P}$
imbalance between the left and right movers, producing a non-zero current 
$\partial E_n / \partial\phi \sim 1/\sqrt{L}$. 

\begin{figure}
\centerline{\epsfxsize=5.3cm \epsffile{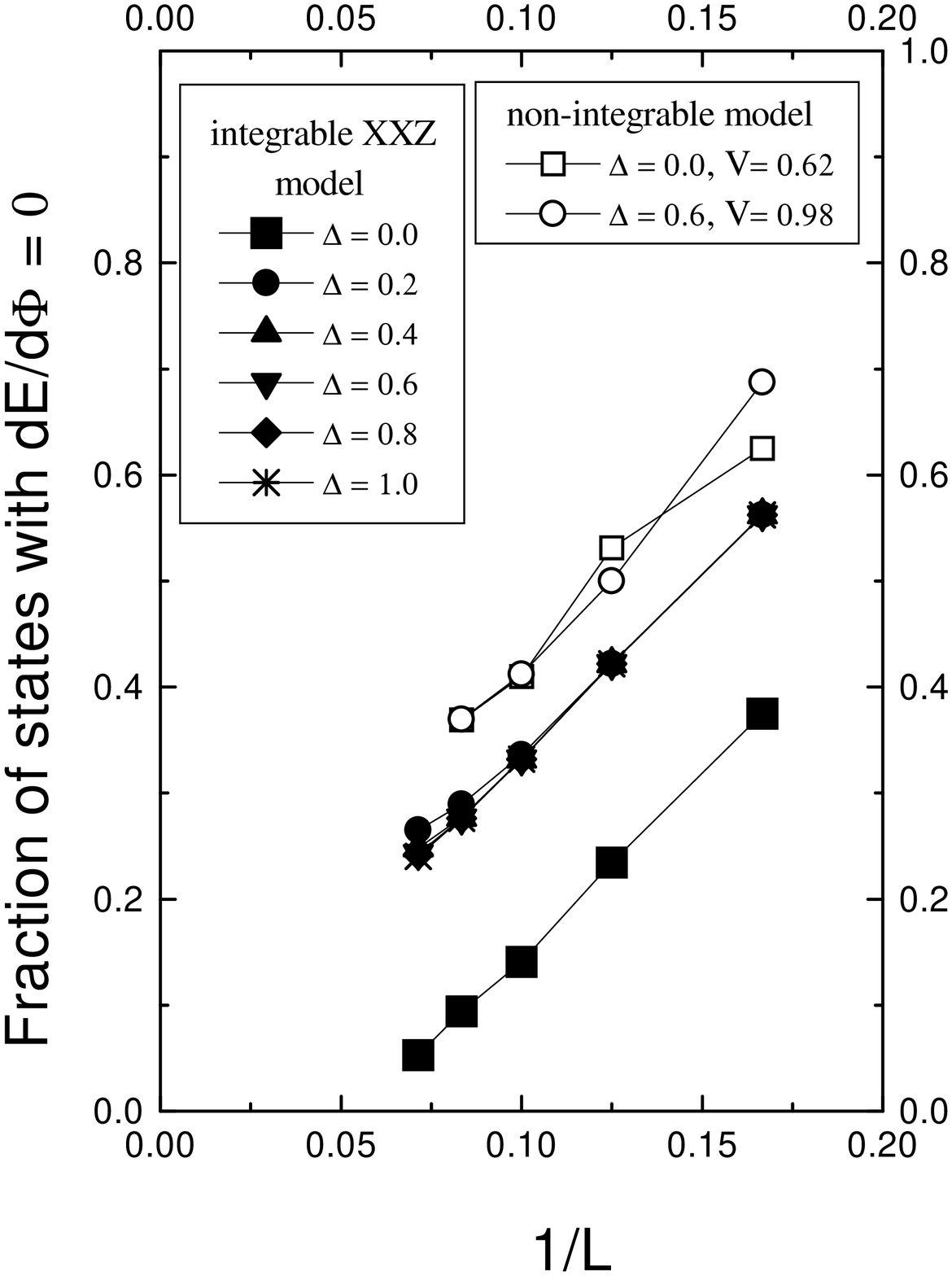}}
Fig 3.
{Fraction of states with zero current} 
\label{fig3}
\end{figure}

The interaction affects current carrying states in two ways. As we noted above these states 
come in degenerate pairs. If the momentum of two degenerate states differs exactly by
a reciprocal lattice vector, the these states will be mixed by the Umklapp interaction term 
which will destroy the degeneracy. Consequently these states will no longer carry current. 
On the other hand if two degenerate states differ by a momentum which is incommensurate
with a reciprocal lattice vector, they cannot be mixed by the Umklapp interaction. The
interaction can mix a given current-carrying state with another  current-carrying states
changing the value of the total current carried. To analyse these effects we plot in Fig. 3
the fraction of states with  $\partial E_n / \partial \phi (\phi\rightarrow 0) = 0$ as
a function of system size for the XXZ model with varying interaction strength $\Delta$
(solid symbols). One sees that adding an interaction sharply increases the fraction of
non-current-carrying states, but this fraction remains small and decreases with system size.

\begin{figure}
\centerline{\epsfxsize=5.3cm \epsffile{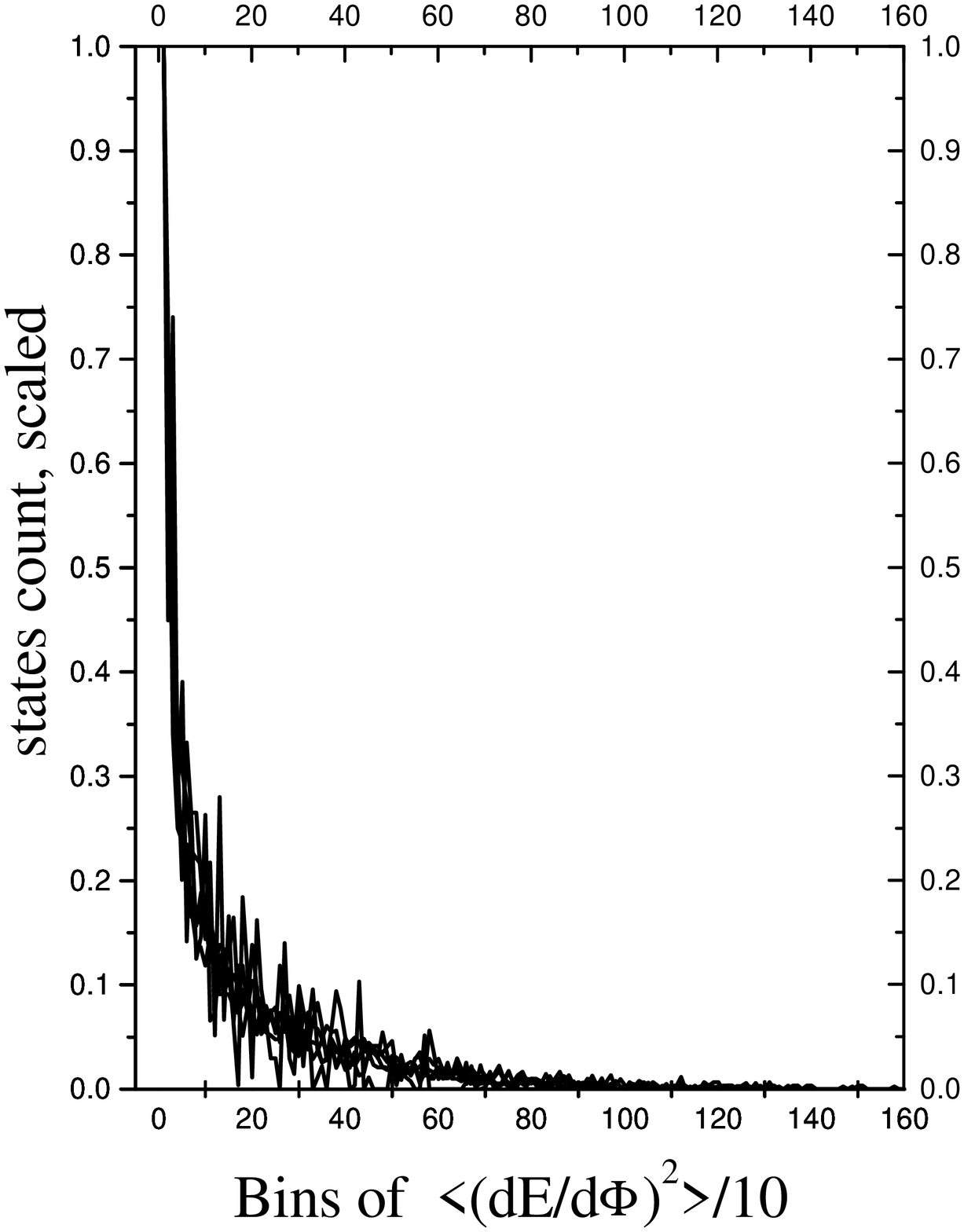}}
Fig 4.
{Histogram of current values; lines for different $\Delta$ are indistinguishable} 
\label{fig4}
\end{figure}

We now consider the statistical distribution of the currents carried by the eigenstates.
We show in Fig. 4 a histogram of $(\partial E_n / \partial \phi)^2$ values for $L=14$ and
all previously considered values of $\Delta$. For each $\Delta$ the $x$-axis has been
scaled so $x=10$ corresponds to $(\partial E_n / \partial \phi)^2$ equal to the average
value. The data have been grouped in to bins of width $0.1$ of the average 
$(\partial E_n / \partial \phi)^2$ and the $y$-axis has been scaled so that $y=1$
represents the number of states in bin 1.With this choice of scaling the distributions
for different $\Delta$ are indistinguishable: the interaction does not change the
shape of the distribution, but merely reduces the avarage value of 
$(\partial E_n / \partial \phi)^2$.

We now consider the effect of an interaction that spoils the integrability of the XXZ 
model by adding the next-nearest neighbor interaction 
$V \sum \limits_{i} S^z_i  S^z_{i+2}$ to the Hamiltonian Eq.\ (\ref{heis}).  As shown
in Fig. 3 this term lifts more degeneracies, so that more states carry zero current.
However the relative change in the fraction of these states is small
and the size dependence is similar to the integrable case.  

The effect of the non-integrable interaction is mostly to reduce the values  of the current
carried by the remaining degenerate states as is illustrated on Fig. 5. At least for large $V$,
$(D_2T)(L\rightarrow\infty) \rightarrow 0$, implying that the current carried by a
typical state, although not $0$, is much less than $1/\sqrt{L}$ and is presumably $o(1/L)$.
Interesting, the effect seems not to occur for $V<\Delta$. For $\Delta = 0$ Fig. 5 shows clearly
that even $V=0.02$ leads to $D_2T$ which vanishes as $L\rightarrow\infty$, whereas for 
$\Delta = 0.2$ large $V=0.62$ leads to a vanishing $(D_2T)(L\rightarrow\infty)$,  while
the effect of $V=0.02$ is much smaller than for $\Delta = 0$ and our data are consistent
with a non-zero $(D_2T)(L\rightarrow\infty)$. Our system sizes are too small to allow us
to make a definite statement about $(D_2T)(L\rightarrow\infty)$ for $V<\Delta$, but clearly
the relative size of the effect of the non-integrable interaction depends strongly on 
the ratio $V/\Delta$.

\begin{figure}
\centerline{\epsfxsize=5.3cm \epsffile{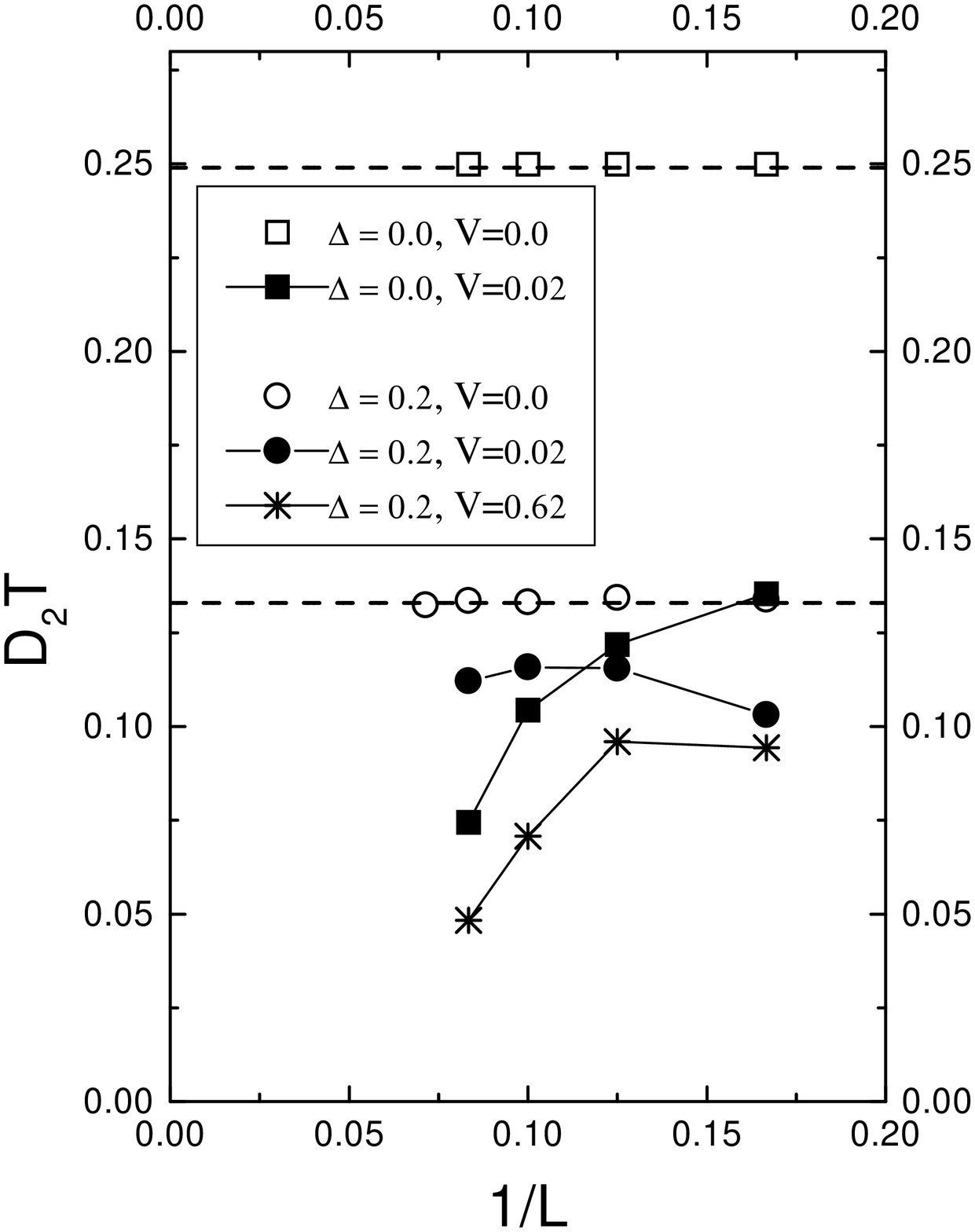}}
Fig 5.
{$D_2 T(1/L)$ comparing integrable and non-integrable cases; the lines are guides to the eye.} 
\label{fig6}
\end{figure}

In conclusion we have studied the spin transport in the Heisenberg model by calculating the
finite temperature stiffness for small system sizes. The data presented in Fig. 2 show that 
for the available sizes $D_2$ is greater than zero and extrapolates to a non-zero value in 
the thermodynamic limit. At small $\Delta$ this extrapolation seems unambiguous in agreement 
with the perturbation theory results \cite{meme}, which are valid for small $\Delta$ where 
the Umklapp is irrelevant in the renormalisation group sence. For larger $\Delta$ the data 
seems to slowly decrease with size extrapolating to some small but non-zero value as 
$1/L \rightarrow 0$. For the isotropic Heisenberg point ($\Delta = 1$) the considered sizes 
are too small to make conclusive predictions about the thermodynamic limit behavior perhaps 
because the Umklapp becomes marginal at this point. 

The non-zero stiffness at $T>0$ means that the model posess a macroscopic fraction of
current carrying states. These states come in degenerate pairs, so the stiffness essentially 
is a measure of non-trivial degeneracy of the energy eigenvalues. We argue that this 
degeneracy is the underlying physical reason for the ballistic transport in the XXZ model.
However, the degeneracy of the XXZ model is due to its integrability and it was earlier argued 
\cite{zot2} that the anomalous transport is due to the large number of conservation laws 
which characterise an integrable model. We believe that among all those conservation laws
there should be only one that is responsible for the degeneracies and therefore for the anomalous 
transport. Based on our results for the non-integrable model with the next-nearest neighbor 
interaction we infer that this conservation law is probably not specific to integrable models. 
The next-nearest neighbor interaction destroys the integrability but still leaves a macroscopic 
fraction of degenerate (current-carrying) states. In the integrable models (except possibly for
$\Delta = 1$) and possibly in the non-integrable models (for $V<\Delta$) the typical current
carried by these states is $\sim 1/\sqrt{L}$ leading to non-zero stiffness as 
$L\rightarrow\infty$.
For $V>\Delta$ we see the typical current is much smaller, leading to 
vanishing stiffness and speculate that there is a length scale $\xi(V,\Delta)$ such
that for system sizes $L < \xi$ the current carried by a typical state
scales as $1/\sqrt{L}$ but that for $L > \xi$ the current scales as
$\xi^{-1/2}/L$.

We are grateful to Lev Ioffe for stimulating our interest in the problem and for helpful
discussions and to Ian Affleck for pointing out an error in the earlier draft of this paper.
A. J. M thanks the Physics Department of Rutgers University for hospitality and NSF DMR 9707701 
for support.

\end{multicols}
\end{document}